\documentclass[preprint2]{aastex}

\usepackage{apjfonts}

\usepackage{graphics}
\usepackage{psfig,natbib}
\usepackage{psbox}

\newcommand{\kms}	{km~s$^{-1}$}
\newcommand{\h}   	{$h^{-1}\,$~kpc}

\newcommand{\lya}	{Ly$\alpha$}

\newcommand{\flux}      {ergs~cm$^{-2}$~s$^{-1}$~\AA$^{-1}$}

\lefthead{D\lya\  from a nearby LSB}
\righthead{D. V. Bowen et al.}

\received{}
\accepted{}

\begin{document}

\title{Damped Lyman-$\alpha$ absorption from a nearby Low Surface
Brightness galaxy\altaffilmark{1,2}}

\altaffiltext{1}{Based on observations with the NASA/ESA
Hubble Space Telescope, obtained at the Space Telescope Science
Institute, which is operated by the Association of Universities for
Research in Astronomy, Inc., under NASA contract NAS5-26555.}
\altaffiltext{2}{Based on observations obtained with the Apache Point
Observatory 3.5~m telescope, which is owned and operated by the
Astrophysical Research Consortium.}

\author{David V.~Bowen, 
Todd M.~Tripp, \& Edward B.~Jenkins}

\affil{ $\:$}

\affil{Princeton University Observatory, Princeton, NJ 08544} 

\begin{abstract}

Ground-based \& HST images of the nearby galaxy SBS~1543+593
($z=0.009$) show it to be a Low Surface Brightness
(LSB) galaxy with a central surface brightness of $\mu_B(0)\:=23.2$
mag arcsec$^{-2}$ and scale length 0.9~\h , values typical for the
local LSB galaxy population.  The galaxy lies directly in front of the
QSO HS~1543+5921 ($z=0.807$); an HST STIS spectrum of the quasar
reveals a damped \lya\ (D\lya ) line at the redshift of the interloper
with an H~I column density of $\log N$(H~I)$\:=\:20.35$, as well as
several low-ionization metal lines with strengths similar to those
found in the Milky Way interstellar medium. Our data show that LSB
galaxies are certainly able to produce the D\lya\ lines seen at higher
redshift, and fuels the speculation that LSB galaxies are a major
contributor to that population of absorbers.

\end{abstract}


                      \section{Introduction}

In the course of spectroscopic follow-up observations of quasar
candidates in the Hamburg Quasar Survey, a redshift of $z=0.807$ was
measured for the QSO HS~1543+5921.  However, Schmidt-plate images
showed the object to be very extended, suggesting that the QSO might
be centered on a low redshift galaxy.  Follow-up observations by
\citet[hereafter RH98]{RH98} found the QSO to be aligned with the
foreground galaxy SBS~1543+593, and measurement of an H~II region in
a spiral arm revealed the galaxy to be at $z=0.009$.

Although detecting QSOs close to nearby galaxies is not difficult,
finding one which shines through the center of a galaxy and which is bright
enough to be observed spectroscopically with HST is extremely rare. In
this paper we present ground-based and HST optical images of
SBS~1543+593 and show that the galaxy is actually a Low Surface
Brightness (LSB) galaxy. We also present an HST spectrum of the
background QSO HS~1543+5921, which reveals a damped \lya\ (D\lya)
absorption line at the redshift of the foreground galaxy.

                    \section{Imaging: Galaxy Properties}

Three 300 sec $R$-band exposures of the QSO-galaxy pair were taken
05 Aug 2000 at the ARC 3.5~m telescope using SPIcam 
at the Apache Point Observatory.
A 2048x2048 SITe CCD was binned on-chip 2x2 to produce
0.28x0.28 arcsec pixels. The data were reduced in the conventional way
and calibrated photometrically with nearby standard stars. Although
conditions were not photometric, calibrators were taken shortly after
images of SBS~1543+593, and are believed to result in a photometric
zero point good to within $0.1-0.2$ mags. 

\begin{figure}[th]
\hspace*{-0.75cm}
\psbox[xsize=0.55\textwidth]{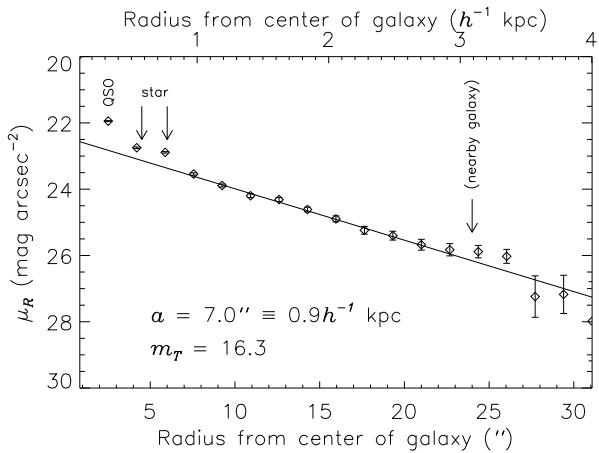}
\caption{\footnotesize{\label{fig_sb} Plot of $R$-band surface brightness for 
SBS~1543+593. Positions of
contributing features are marked. A theoretical profile of 
$\mu_R\:=\:22.6+1.086(r/7.0)$ is also drawn.}}
\end{figure}

The coadded data resulted in an image similar to that shown in RH98,
and are not reproduced here. We used these data, however, to produce a
surface brightness profile of the galaxy, shown in
Figure~\ref{fig_sb}. The profile near the center of the galaxy is
obviously contaminated by  flux from the QSO and the bright star
identified by RH98, but the remaining points can be well fit with a
standard exponential profile, $\mu_R\:=\:\mu_R(0)+1.086(r/a)$, where
$\mu_R(0)$ is the central surface brightness in mag arcsec$^{-2}$, $r$
is the radius from the galaxy's center, and $a$ is the scale
length. The fit to the data in Figure~\ref{fig_sb} shows that the
galaxy has $\mu_R(0)=22.6$ mag arcsec$^{-2}$ and a scale length of 7
arcsec, or 0.9~\h\ at the redshift of the
galaxy.\footnote{$h\:=\:H_0/100$, where $H_0$ is the Hubble constant,
and $q_0\:=\:0$ is assumed throughout this paper} We make no
correction to $\mu_R(0)$ for galaxy inclination since the optical
depth effects in LSB galaxies are not known, and may be less
significant than in High Surface Brightness (HSB) galaxies if the dust
content in LSB galaxies is lower.  If dust does play a significant
role in LSB galaxies, then $\mu_R(0)$ is an upper limit and the galaxy
may be fainter.  The total $R$-band magnitude of
the galaxy, $m_R(T)$ can be derived from the fit to the surface
brightness profile, $m_R(T)\:=\:\mu_R(0) - 2.5\log (2\pi
a^2)\:=\:16.3$, in good agreement with a simple integration of the
counts which excludes star and QSO. This corresponds to an
absolute magnitude of $M_R - 5\log h\:=\:-15.9$.

We can compare the properties of this galaxy with those found for
other local LSB galaxies by correcting the surface brightness in the observed
$R$-band to that in the $B$-band. \citet{deBlok96} find that for a
sample of nearby LSBs, $\mu_B(0)\:=\:\mu_R(0) + 0.78$, which gives
$\mu_B(0)\:=23.2$ mag arcsec$^{-2}$ for SBS~1543+593. Compared to the
sample studied by \citet{deBlok95}, SBS~1543+593 is a regular
LSB galaxy, with $\mu_B(0)$ and $h$ close to the median values of
$\mu_B(0)=23.4$ and $a=3.2$~\h\ for local LSB galaxies.

\begin{figure*}
\centerline{\psfig
{figure=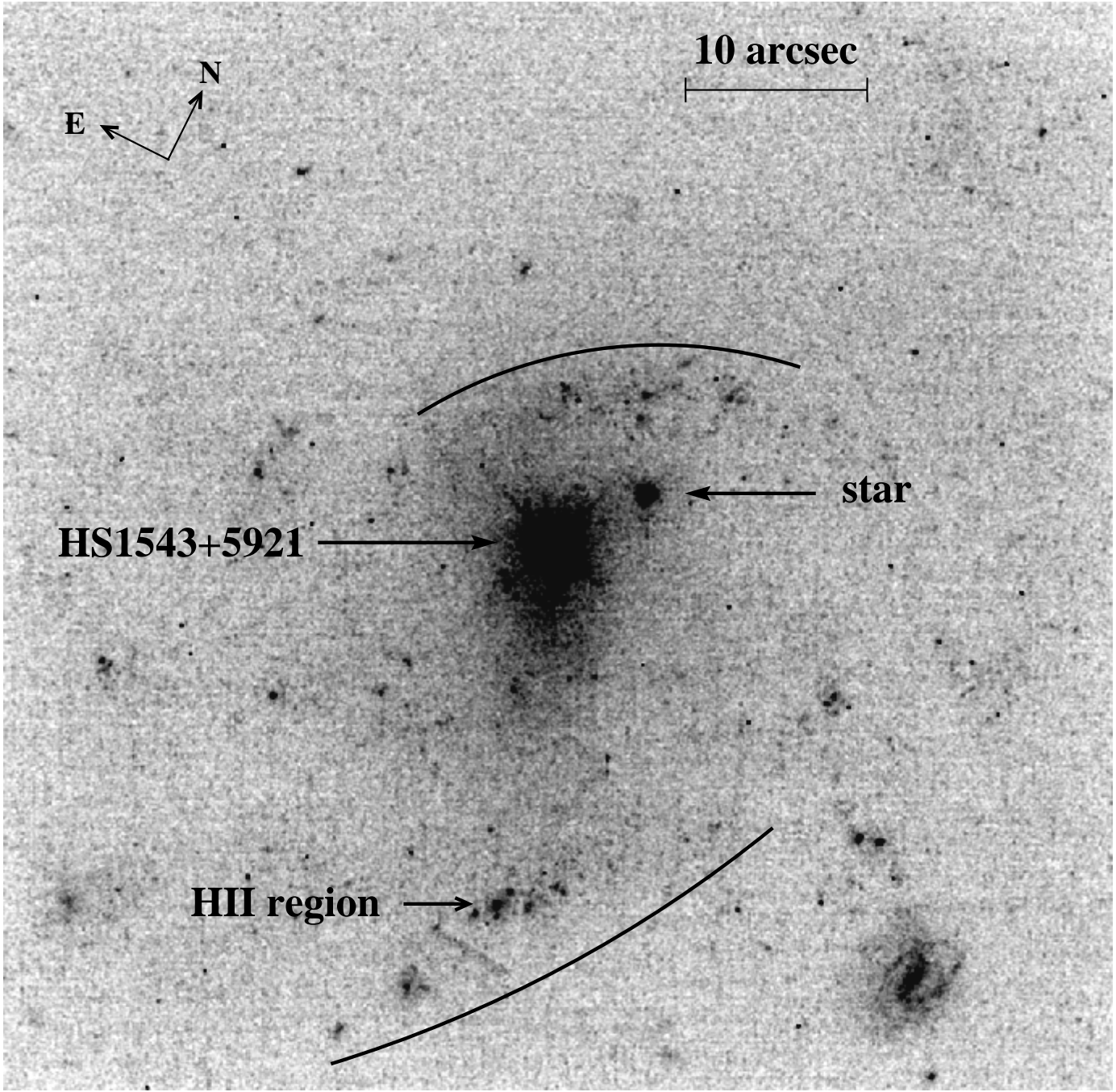,height=18cm,angle=0}}
\caption{\label{fig_stis}Reproduction of a 804 sec coadded 
HST STIS image of SBS~1543+593 taken
with no filter. The H~II region whose redshift was measured by RH98 is
marked. Since the low surface brightness features are difficult to
reproduce, we have delineated the outer edges of the spiral arms 
to guide the eye.}
\end{figure*}

Three HST STIS images were taken of the QSO-galaxy pair on 18 Sep 2000
with the CCD detector and no filter.  The three
exposures were medianed to remove cosmic rays, and hot pixels were
identified and set to a local background. The coadded image represents
a total exposure time of 802 sec and is reproduced in
Figure~\ref{fig_stis}, with the data smoothed by a boxcar filter of
size 3x3 pixels to show the LSB features better.  The field of view is
$\sim 50$x50 arcsec and the throughput of the detector is such that
the image records photons with wavelengths between 2000 and 11,000~\AA
, with the sensitivity peaking at 5852~\AA . The image shows that
SBS~1543+593 is composed of two primary spiral arms, largely defined by
H~II regions within the arms. In Figure~\ref{fig_stis} we mark the
outer regions of these two arms with curves. The H~II region observed
spectroscopically by RH98 is marked, and is seen to break up into several
discrete regions. There may be evidence for additional spiral
structure in the image; for example, a faint diffuse region to the left of
the ``H~II region'' label at the bottom left of  Figure~\ref{fig_stis}
may itself be an H~II region and mark the end of a faint spiral arm
which winds back towards the more prominent southern H~II cluster
already discussed. A second galaxy bottom right of the figure,
which can be seen in RH98's image, and identified by them as a point
source, is here easily resolved into a barred
spiral. No redshift is measured for this galaxy, but given the small
angular size and high surface brightness, it is probably background to
SBS~1543+593.

\section{STIS spectroscopy: damped \lya\ from SBS~1543+593}

\begin{figure*}[th]
\hspace*{1.4cm}
\psbox[xsize=1.0\textwidth,rotate=l]{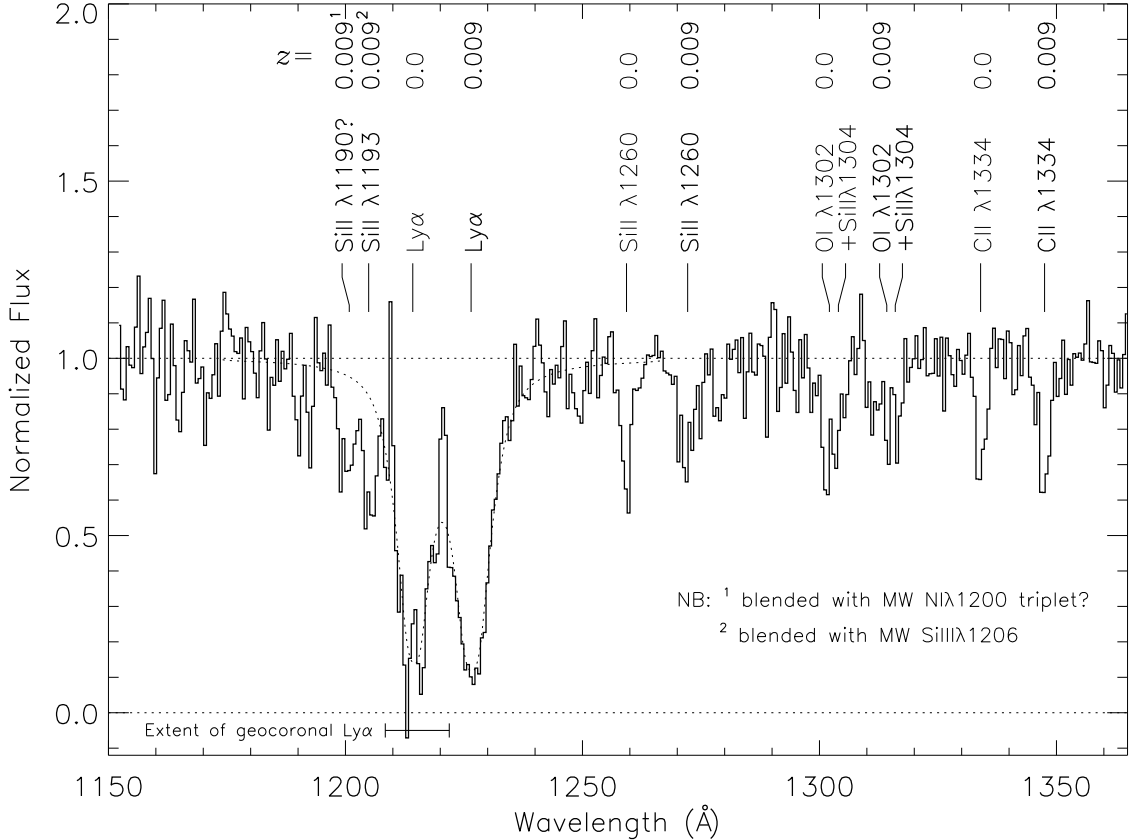}
\caption{{\footnotesize \label{fig_spec} Normalized 2139 sec 
coadded G140L STIS spectrum of HS~1543+5921. 
The \lya\ lines from our own galaxy and from SBS~1543+593 at $z=0.009$
are detected, along with several low-ionization species. Redshifts of
the lines are shown above their identification. Theoretical \lya\ line
profiles are overplotted, corresponding to $\log N$(H~I)$=20.26$ from
the Milky Way and 20.35 from SBS~1543+593. The flux at the bottom of
the \lya\ absorption lines do not reach zero due to the shape of the G140L Line
Spread Function, which has broad, extended wings (see
text). Also labelled is the minimum and maximum extent of the
geocoronal \lya\ emission which was subtracted from the data to
produce this spectrum.
}}
\end{figure*}

Three STIS FUV-MAMA spectra were obtained 18 Sep 2000 with the G140L
grating and 52x0.5 aperture for a total exposure time of 2139 sec (HST
Archive dataset o5l301010).  To improve the signal-to-noise of the
spectrum over that derived from the pipeline calibration, we used the
IRAF {\tt apall} routine to extract the individual exposures from the
x2d file, paying particular attention to the subtraction of geocoronal
\lya\ and O~I~$\lambda 1302$ emission which filled the aperture.  The
co-added data were normalized using a fifth-order cubic spline, and a
portion of the resulting spectrum is shown in Figure~\ref{fig_spec}.
Not shown in that figure is a broad emission line centered at
1397.8~\AA , which we identify as Ne~VIII~$\lambda\lambda 770,780$
from the QSO, giving an emission redshift of $z=0.804$.  The value of
the flux at 1220~\AA\ is $2.6\times 10^{-15}$ \flux .

Absorption lines in the spectra were identified, and equivalent widths
measured, using standard procedures, and lines found arising from
SBS~1543+593 are listed in Table~\ref{tab_lines}. \lya\ is clearly
seen from the galaxy at precisely the redshift expected, and appears
to be slightly stronger than Galactic \lya. Strong lines of
Si~II~$\lambda 1260$, O~I~$\lambda 1302$+Si~II~$\lambda 1304$, and
C~II~$\lambda 1334$ are detected at strengths similar to Milky Way
absorption lines.  Although the region blueward of \lya\ is confused
at this resolution, it is likely that we also detect Si~II~$\lambda
1193$ blended with Galactic Si~II~$\lambda 1206$. We may also detect
Si~II~$\lambda 1190$ at 1200.8~\AA, but,
compared to the other Si~II lines,
the absorption may be too strong
to be from the 1190~\AA\ line
alone. Although the Milky Way N~I~$\lambda 1200$ triplet is expected
to be present, the predicted wavelength of the line does not match
this feature well.  Interestingly, with Galactic co-ordinates of
$l=92.4$ and $b=46.4$, this sightline passes through the middle of
High Velocity Cloud (HVC) Complex C \citep{wakk91}. However, for the
two lines at 1200.8 and 1204.9~\AA\ to be identified as \lya\ from the
HVC Complex, they would have to have velocities of $\sim -3700$ and
$-2600$~\kms\ respectively, velocities much too large to be associated
with HVCs.  There are no other obvious absorption line systems which
can be identified in the spectrum that could give rise to metal lines
at the wavelength of the 1200.8~\AA\ feature, so its identification
remains ambiguous.

Measurement of the neutral hydrogen column density along the line of
sight is complicated by the shape of the Line Spread Function (LSF)
generated by the G140L grating and 52x0.5 aperture.  The resolving
power of the G140L is measured to be between 200$-$300~\kms\
\citep{kimble98}, but the LSF is not gaussian, and is composed instead
of a central narrow core with broad, extended wings.  Although
the widths of the \lya\ absorption lines from both the Milky Way and
SBS~1543+593 are much greater than the FWHM of the core of the LSF, the
inclusion of extended wings means that part of the absorption profile
is over a velocity interval at least as large as the intrinsic widths
of the \lya\ lines. The result is that the core of the lines do not
reach a flux of zero as might be expected for damped lines if the LSF
were a gaussian only a few hundred \kms\ wide.

Hence, to model the \lya\ lines from the Galaxy and from SBS~1543+593,
we have convolved theoretical line profiles with the available
LSF\footnote{Available from the STIS web pages at www.stsci.edu}
calculated at 1200~\AA .  The H~I column density from our own Milky
Way is known to be $\log N$(H~I)$=20.26$ from, e.g., the Bell Labs H~I
Survey \citep{BellHI}\footnote{Obtained from
asc.harvard.edu/toolkit/colden.jsp}. The signal-to-noise of the
Galactic \lya\ absorption line is poor, due to the subtraction of
intense geocoronal \lya\ emission; the minimum and maximum extent of
the subtracted profile is shown bottom of Figure~\ref{fig_spec},
demonstrating that the spurious point at 1209.4~\AA, and the pixels
separating the Galactic and extragalactic \lya\ which seem anomalously
high, are a result of the subtraction of the geocoronal \lya .  Hence
we fitted theoretical line profiles to the \lya\ absorption lines
keeping $\log N$(H~I) fixed for the Milky Way absorption, while
allowing the redshift, column density and Doppler parameter to vary
for absorption from SBS~1543+593.  The resulting fits are shown in
Figure~\ref{fig_spec}. The profile for Milky Way absorption well fits
the data, despite using a value of $N$(H~I) known a priori, and for
SBS~1543+593 we derive $\log N$(H~I)$=20.35$. We note that by {\it
not} using an LSF with extended wings, assuming a simple gaussian LSF,
and correcting the baseline of the spectrum so that the cores of the
\lya\ lines reached zero flux, $N$(H~I) would be over-estimated by
$\sim 0.1$ dex.  A value of $\log N$(H~I)$=20.35$ is consistent with
that derived from simply calculating $N$(H~I) from the measured
equivalent width assuming a damped line: for $W_\lambda$(\lya
)$\geq\:9.15$~\AA\ (a lower limit since part of the line is lost
though blending with Milky Way \lya ) we derive $\log N$(H~I)$\geq
20.27$.

Errors on the derived value of $N$(H~I) are hard to quantify
precisely, due to uncertain systematic errors, but providing the
line is damped, $N$(H~I) is probably accurate to within 0.1 dex.  It
seems likely, therefore, that SBS~1543+593 is responsible for a D\lya\
line. We note, however, that at the resolution of these observations,
it is still possible that the line could be composed of several
individual components whose distribution of column densities and
Doppler parameters with velocity mimics a single D\lya\ profile.
This would mean that the total $N$(H~I) could be less than the derived
$\log N$(H~I)$=20.35$. If true, however, the similarities in
equivalent width between the Milky Way metal absorption lines and
those arising in the LSB galaxy would imply that the extragalactic
interstellar gas was of a higher metallicity than that intercepted
locally.  Higher resolution observations would
confirm the damped nature of the \lya\ line.

\begin{small}
\begin{table}[hb]
\caption{Lines at $z=0.009$ from SBS~1543+593 \label{tab_lines}}
%\begin{center}
\begin{tabular}{clcl}
\hline\hline
$\lambda$      &  $W_\lambda$    & $\sigma(W_\lambda)$     &       \\
(\AA )         & (\AA )  & (\AA )          & ID    \\
\hline
   1200.8     & 1.60$^1$      & 0.23  & Si~II~$\lambda1190$?  \\
              &               &       & (+N~I~$\lambda1200$ at $z=0$?) \\
   1204.9     & 1.78$^1$      & 0.22  & Si~II~$\lambda1193$ \\
              &               &       & + Si~II~$\lambda1206$ at $z=0$ \\
   1226.5     & 9.15$^1$      & 0.32  & \lya \\
   1272.2     & 1.07          & 0.18  & Si~II~$\lambda1260$ \\
   1314.2     & 1.30          & 0.15  & O~I~$\lambda1302$+Si~II~$\lambda1304$   \\
   1347.5     & 1.19          & 0.12  & C~II~$\lambda1334$ \\
\hline
\multicolumn{4}{l}{$^1$ Lower limits---blended with Milky Way lines}\\
\end{tabular}
%\end{center}
\end{table}
\end{small}

\section{Discussion}

Our optical and spectroscopic data have shown that SBS~1543+593 is a
LSB galaxy causing a D\lya\ system at $z=0.009$ in the spectrum of
HS~1543+5921. This makes it the lowest redshift D\lya\ system
discovered outside of the local group. The identification of D\lya\
from a {\it known} LSB is important, for the following reasons.  So
far, results from ground-based and HST imaging of fields
around QSOs known to show $z<1$ D\lya\ lines have been surprising, with
the detection of a whole variety of galaxy types, including normal
early and late-type HSB spirals, and amorphous LSB galaxies,
identified as responsible for the absorption \citep{steidel95,
steidel94, Lanz97, LeBr97, RaTu98, pett00}.  This wide variety of
absorber types has led to the speculation that D\lya\ systems may
not simply signal the presence of normal gas-rich spiral galaxies
after all, as has been postulated since their discovery
\citep{wolfe86}. It is important to note, however, that in most cases
there exists no redshift information for the purported absorbers.
Proximity to the line of sight is no guarantee that an `identified'
object is the absorber, since there are often several absorption
systems at redshifts other than that of the D\lya\ line along the QSO
line of sight. If galaxies are responsible for these other systems,
then the chance of mis-identification is high (particularly if
D\lya\ systems do not arise in normal galaxies).  Hence the detection
of a D\lya\ line from SBS~1543+593 is unique, in that  we know
unequivocally that the absorber is an LSB galaxy.

It also seems likely that if SBS~1543+593 were moved to a redshift
similar to those of the $z<1$ D\lya\ systems already studied, it would
be extremely difficult to detect, partly due to its low surface
brightness and partly due to its close proximity to the QSO and its small
angular size. In fact, in this case, it is more likely that the nearby
barred spiral galaxy south-west of the pair, which we take to be
background to SBS~1543+593, would be identified as the D\lya\ absorber
if no redshift information were available.

Our observations of SBS~1543+593 support the idea that LSB galaxies
may contribute significantly to the population of D\lya\ absorbers, as
initially suggested by \citet{Imp89}. Although finding a D\lya\ from
an LSB galaxy does not prove that {\it all} D\lya\ systems are LSB
galaxies, our detection does prove that LSB galaxies can produce such
systems. 

Perhaps more significant is the detection of this QSO-galaxy
pair in the first place. Finding any bright QSO shining through the
center of a nearby galaxy is extremely rare, and it is intriguing that
an LSB galaxy is the interloper.  LSB galaxies are believed to be
relatively free of dust, and hence optically thin to the photons of
background objects --- these may therefore be the best types of
galaxies to allow the light of quasars to shine through. This
potential selection effect has been a concern for interpreting the
copious abundance measurements of high-redshift D\lya\ systems: if a
particular type of galaxy preferentially favors QSO light passing
through it, then the derived metallicities are only applicable for
that type of galaxy.  
There are also theoretical bases for the idea
that LSB galaxies hold most of the high column density H~I
cross-section: for example, \citet{raul99} predicted that HSB disks
consume neutral gas too fast to explain the observed evolution in the
neutral gas mass density with redshift, and that consumption of
hydrogen by LSB galaxies better fits the abundance measurements.  

The
mere existence of the QSO-galaxy pair discussed herein seems to strongly
support these views. Unfortunately, just 
how common such QSO/LSB-galaxy alignments are at
higher redshift will always be difficult to determine,
due to the intrinsic faintness of the galaxies, especially if the
galaxy is close to the QSO sightline as with this pair.
Finally, we note that HS~1543+5921 is bright enough to be used to
measure abundances in the interstellar medium of SBS~1543+593 with
further HST observations. The current data do not have sufficient
resolution for reliable metallicity estimates, so follow-up UV
observations would be extremely valuable. Comparison of gas
metallicities from a known LSB galaxy with higher redshift D\lya\
systems would be useful in deciding if LSB galaxies are, in general,
responsible for the absorption. Measurement of the alpha-to-iron
elemental abundances in SBS~1543+593 could also be used to derive the
star-formation history of the galaxy, and relative metal abundances
may provide insight into its dust content. Lastly, comparison of the
kinematic structure of metal lines arising from SBS~1543+593 could
be matched to global H~I kinematics derived from 21~cm emission
maps of the galaxy, and to profiles seen in high-redshift D\lya\
lines which are believed by some authors to be indicative of rotating disks.

\acknowledgments

Support for this work was provided in part from NASA through grant
NAS5-30110.

\bibliographystyle{apj}
\bibliography{apj-jour,bib1}

\end{document}